\documentclass[info,debug,overfull]{epl}%
\usepackage{graphicx}
\usepackage{amsmath}
\usepackage{amsfonts}
\usepackage{amssymb}%
\institute{
\inst{1}
Department of Mechanical Systems Engineering, Toyama Prefectural University,
Kosugi-machi, Toyama 939-0398, Japan\\
\inst{2}
Department of Physics, Chuo University, Bunkyo-ku, Tokyo 112-8521, Japan\\
\inst{3}
Department of Physics, Graduate School of Humanities and Science,
Ochanomizu University, Otsuka, Bunkyo-ku, Tokyo 112-8610, Japan\\
\inst{4}
Laboratoire de Physique de la Mati\`ere Condens\'ee, Coll\`{e}ge de France,
75231 Paris Cedex 05, France}
\begin{document}

\title{Bouncing gel balls: impact of soft gels onto rigid surface }
\author{Yoshimi Tanaka,\inst{1} Yoshihiro Yamazaki\inst{2} and Ko Okumura\inst{3,4}}
\maketitle

\begin{abstract}
After thrown onto a solid substrate, very soft spherical gels bounce
repeatedly. Separate rheological measurements suggest that these balls can be
treated as nearly elastic. The Hertz contact deformation expected in the
static (elastic) limit was observed only at very small impact velocities. For
larger velocities, the gel ball deformed into flattened forms like a pancake.
We measured the size of the gel balls at the maximal deformation and the
contact time as a function of velocities for the samples different in the
original spherical radius and the Young modulus. The experimental results
revealed a number of scaling relations. To interpret these relations, we
developed scaling arguments to propose a physical picture.

\end{abstract}

\shorttitle{Bouncing gel balls} \shortauthor{Y. Tanaka, Y. Yamazaki and K. Okumura}

\section{Introduction}

Impact of spherical objects is concerned with a wide range of subjects from
daily life to solid mechanics \cite{Johnson,Goldsmit,Stronge}. Accordingly, it
has been studied for a long time in different contexts and still continues to
be an important research subject. The Hertz contact theory \cite{Hertz} is a
classic example \textit{in the static limit} within the linear elastic theory.
For larger impact velocities, internal vibrations\cite{Gerl} and dissipative
processes such as solid viscosity \cite{Kuwabara}, plastic
deformation\cite{Goldsmit,Labous} should be considered. Impact of solid
spheres in a granular medium has been found to be important fundamental
processes in granular flow, and has attracted a considerable attention
\cite{Labous,Morgado,Hayakawa}. Impacts of balls of soft matter could also be
an interesting problem: unique phenomena are expected due to a variety of
constitutive equations and due to large deformations. For example, impact or
dynamics of liquid balls has attracted a wide audience
\cite{Vance,David,RQ2000,MP,Vance2,Brechet,Pascale,Maha}.

In this paper, we study another example from soft matter: impacts of soft gel
balls vertically impinging onto a solid substrate. There the Hertz contact
deformation is observed only for small velocities. For larger velocities, gel
balls are flattened globally during the impact on the substrate, i.e.,
expanded into the lateral direction. We present experimental data on the
lateral dimension of balls at the maximal deformation and the contact time as
a function of impact velocities to show the existence of a number of scaling
relations. These relations are interpreted via a naive physical picture.

\section{Sample gels}

\label{Exp}

We used a series of acrylamide gels with the same polymer concentration but
with different cross-link densities; acrylamide monomer (AA, $M_{w}$ =71.08)
constitutes sub-chains while methylenebisacrylamide (BIS, $M_{w}$ =154.17)
cross-links. The amount of each reagent for preparing acrylamide gels is shown
in Table 1. Ammonium persulphate (1wt \% of AA) and
tetramethylethlylenediamine (0.25vol \% of water) were added to initiate and
to accelerate the radical polymerization of AA and BIS. The pre-gel solutions
were sealed into a spherical mold ($R=$14mm or 31.5mm) consisting of two
hemispherical shells. Gelation reaction continued for 24 hours at 30 $^{\circ
}$C.

Rheological characterization was carried out on cylindrical gels with a
rheometer (REOGEL, UBM Co.) in an oscillatory compression mode. As a result,
we found that the mechanical responses of our samples can be all regarded as
\emph{nearly elastic}. The values of $\tan\delta$ are less than 0.02 and the
real part of the complex modulus $E^{\prime}$ is almost constant in the range
of strain frequency between 0.1Hz and 100Hz for all samples except BIS4 (Here,
$\delta$ is the phase difference between the stress and the strain). Even for
this softest sample, $\tan\delta$ increases by a fairly small amount with $f$
(for example, $\tan\delta\simeq0.02$ at $f=0.01$Hz, $\tan\delta\simeq0.05$ at
$f=50$Hz and $\tan\delta\simeq0.1$ at $f=100$Hz). Table \ref{table1} also
shows the modulus $E^{\prime}$ at $f=10$Hz that is regarded as \emph{the
static Young's modulus} $E$ in the following.\begin{table}[ptb]
\begin{center}%
\begin{tabular}
[c]{c||ccc|c}\hline\hline
sample & Water & AA & BIS & $E$(Pa)\\\hline
BIS4 & 100cc & 10g & 0.04g & 1.24$\times$10$^{4}$\\
BIS10 & 100cc & 10g & 0.1g & 2.71$\times$10$^{4}$\\
BIS15 & 100cc & 10g & 0.15g & 3.87$\times$10$^{4}$\\
BIS20 & 100cc & 10g & 0.20g & 4.56$\times$10$^{4}$\\\hline\hline
\end{tabular}
\end{center}
\caption[table]{The composition of four samples of acrylamide gel and their
Young moduli $E$.}%
\label{table1}%
\end{table}

\section{Impact experiment}

Gel balls freely fall on a fixed aluminum plate of 20mm thickness. Before the
impact, the gel ball is pinned at a height $L$ from the plate by a tube
sucking air weakly. The gel ball begins to fall by switching-off the sucking,
and then impinging onto the aluminium substrate. We coated the surface of
balls in white with aluminum oxide powder to avoid sticking of the gel balls
to the substrate (this is especially important for low impact velocities). The
impact velocity $V$ can be determined from the relation $V=\sqrt{2gL}$. The
impact processes are recorded by a high-speed CCD video camera (Motion Coder
Analyzer SR: Kodak Co.) with recording rates of 1000 FPS (samples of
$R=31.5$mm) or 2000 FPS (samples of $R=14$mm).

\section{Experimental Results}

\begin{figure}[tbh]
\includegraphics[height=2.8cm]{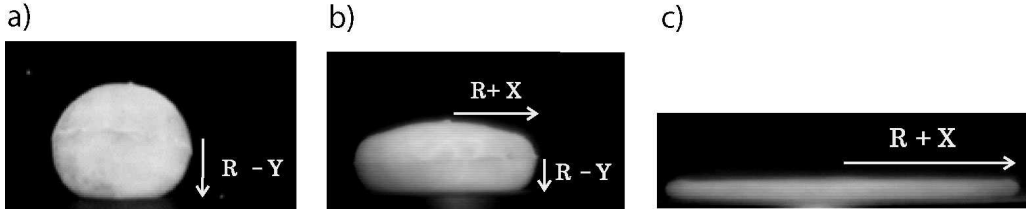}\caption{Maximal deformations of gel
balls ($R=31.5$mm). a) Hertz regime for very low impact velocities (BIS10,
$V\lesssim0.4$m/s). b) quasi-ellipsoid regime for intermediate velocities
(BIS4, $V=2.0$m/s). c) pancake regime for high velocities (BIS4, $V=7.0$m/s).
}%
\label{f1}%
\end{figure}We found experimentally that modes of deformation are rather
different depending on the velocities. The form of maximal deformations can be
categorized in three classes: (1) the Hertz type for very small impact
velocities (Fig. \ref{f1}a), (2) quasi-ellipsoid for intermediate velocities
(Fig. \ref{f1}b), and pancake for large velocities (Fig. \ref{f1}c).

For small impact velocities the Hertz-type deformation is theoretically
expected: in the static limit, because of the nearly pure elasticity of gel
balls, the localized Hertz deformation should be observed. However, this
theory should break down for large impact velocities where a ball deforms
non-locally. Indeed, the Hertz regime can be observed only for very small
velocities: the impact in Fig. \ref{f1}a was achieved by dropping the gel ball
from a low height $L$ less than 1cm (without using the sucking system) where a
precise determination of $V$ is difficult ($V\lesssim0.4$m/s) (We note here
that all the data points used in the following plots are limited to the case
where velocity determination can be done rather precisely). For moderate
impact velocities deformation is no longer localized around the contact area
and the ball takes a flattened shape which is quasi-ellipsoidal (Fig.
\ref{f1}b). For large impact velocities the ball is strongly deformed to take
a pancake shape (Fig. \ref{f1}c).

We measured the horizontal radius $R+X$ (see Fig. \ref{f1}) at the maximal
deformation and the contact time at various conditions. As shown below, we
found that these experimental data can be well characterized by the length
scale $R$, the sound velocity $V_{c}$ and the corresponding time scale
$\tau_{c}$:%
\begin{equation}
V_{c}=\sqrt{E/\rho} \label{vc}%
\end{equation}%
\begin{equation}
\tau_{c}=R/V_{c}, \label{tc}%
\end{equation}
where $\rho$ is the density of gel balls ($\rho\simeq1.05\times10^{3}%
$kg/m$^{3}$). Characteristic scales $V_{c}$ and $\tau_{c}$ come out naturally
from theoretical considerations presented below.

\begin{figure}[tbh]
\includegraphics[scale=0.5]{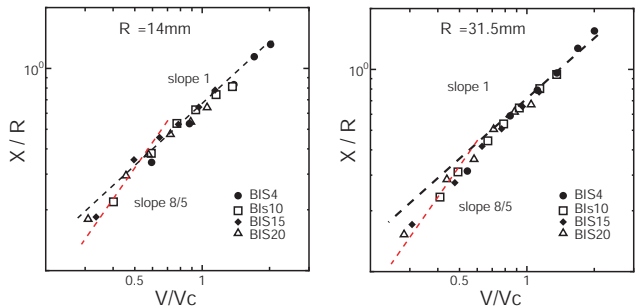}\caption{Maximal deformation $X/R$ as a
function of impact velocities $V/V_{c}$ for (a) smaller samples ($R=$14mm) and
(b) larger samples ($R=$31.5mm). Two dashed lines in the plots corresponds to
lines with slopes 1 and 8/5, respectively. }%
\label{f2}%
\end{figure}Fig. \ref{f2} shows the maximal deformation $X/R$ as a function of
reduced velocities $V/V_{c}$. As mentioned above, owing to the characteristic
scales $R$ and $V_{c}$, the data from the four samples for each size ($R=14$mm
or 31.5mm) collapse well onto a single behavior which can be divided in two
regimes as indicated by two dashed lines with different slopes (8/5 and 1).
For large velocities ($V\gtrsim V_{c}$) it scales as $X\sim V$ while for small
velocities ($V\gtrsim V_{c}$) $X\sim V^{n}$ with $n$ being a certain value
larger than the unity. A theoretical value of the exponent, $n=8/5$, for the
small velocity region seems consistent with the data (see below).

\begin{figure}[tbh]
\includegraphics[scale=0.55]{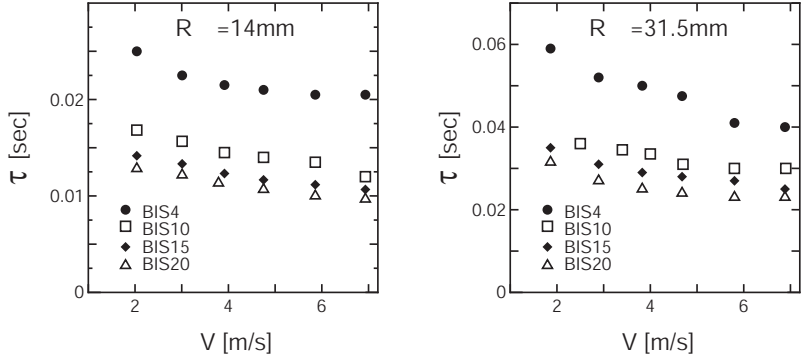}\caption{Contact $\tau$ as a function of
velocities $V$ for gel balls of radius $R=$14mm (a) and $R=$31.5mm (b).}%
\label{f3}%
\end{figure}\begin{figure}[tbhtbh]
\includegraphics[scale=0.55]{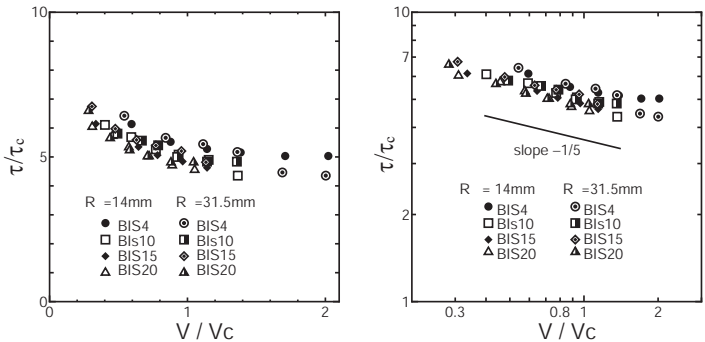}\caption{Reduced plot of a contact time
$\tau$ versus impact velocity $V$. The right plot suggests a scaling relation
$\tau\sim V^{-1/5}$.}%
\label{f4}%
\end{figure}Fig. \ref{f3} shows a contact time $\tau$ as a function of
velocities. The two plots collapse well onto a single master curve in Fig.
\ref{f4}, thanks to the characteristic scales $\tau_{c}$ and $V_{c}$ as
mentioned above. The contact time $\tau$ decreases with increase in
velocities, following a scaling law ($\tau\sim V^{-1/5}$) and seems to
approach a constant value (Figs. \ref{f3} and \ref{f4}). The plateau value is
of the order of millisecond. This value increases with the size $R$ and
decreases with the modulus $E$. Accordingly, the data from different sizes and
moduli collapse by the characteristic scale $\tau_{c}$.

As seen above, both $X$ and $\tau$ can be divided in two regimes by a
characteristic velocity of the order of m/s. This velocity scale increases
with a modulus $E$ (this can be confirmed due to the data collapse via
renormalization of velocity by $V_{c}$).

\section{Theoretical considerations}

As mentioned above and as seen from the snapshots at the maximal deformation,
modes of deformation seem to change with increase in impact velocity: from (1)
Hertz type, to (2) ellipsoid, and then to (3) pancake (see Fig.\ref{f1}). We
shall develop a theory based on these observed forms of deformation. We note
in advance that experimental results can be understood by combining only the
two limiting regimes (Hertz and pancake). However, we still present the result
based on the ellipsoid approximation as reference, which is consistent with
the experimental results in the turn-over regime. We start from scaling
arguments for the Hertz regime only to reproduce the essential properties of
the well-known complete analytical solution. The arguments are then extended
to the other regimes.

In the Hertz regime at small impact velocities, the elastic energy stored at
the maximal deformation can be estimated dimensionally as $U_{H}\sim E\left(
Y/a\right)  ^{2}a^{3}$ with $E$ being the elastic modulus of the ball ($Y$ is
defined as in Fig.\ref{f1} and $a$ is the radius of the contact area); a
strain field of the order of $Y/a$ is localized in a volume of the order of
$a^{3}$ (the deformation field relaxes out at a distance of the order of $a$
from the contact surface because the strain field is governed by the Laplace
equation). Due to the geometrical relation $Y\sim a^{2}/R$ with $R$ being the
radius of the ball, the energy reduces to the well-known form:
\begin{equation}
U_{H}\sim E\sqrt{R}Y^{5/2} \label{UH}%
\end{equation}
Assumption of the energy conservation between the initial time and the
maximally-deformed moment $MV^{2}\sim U_{H}$ ($M\sim\rho R^{3}$ is the mass of
the ball) leads us to an estimate for the contact time $\tau\sim Y/V$, which
results in the well-known relation,%
\begin{equation}
\tau/\tau_{c}\sim(V/V_{c})^{-1/5} \label{tH}%
\end{equation}
where the characteristic scales $V_{c}$ and $\tau_{c}$ given in Eqs.
(\ref{vc}) and (\ref{tc}) have naturally come out as announced. This implies a
moderate increase of the contact time with decrease in velocity. To obtain a
non-trivial $X-V$ relation, we assume that in our "Hertz regime" the shape is
\emph{a sphere cut out by a plane} (see below) and require the
incompressibility condition: the part of the sphere cut out by the substrate
is compensated by an increase in radius, i.e.,
\begin{equation}
a^{2}Y\sim R^{2}X. \label{eqA}%
\end{equation}
Combined with the previous relation of the energy conservation, we have the
relation between impact velocity and deformation in the horizontal direction
as mentioned above:%
\begin{equation}
X/R\sim(V/V_{c})^{8/5} \label{xH}%
\end{equation}

In the ellipsoid regime at intermediate velocities, the maximal elastic energy
is given by $U_{E}\sim E\left(  Y/R\right)  ^{2}R^{3}$; a strain field of the
order of $Y/R\sim X/R$ is distributed within the whole volume of the order of
$R^{3}$. Note here the relation, $Y=2X$, which expresses the condition of
volume conservation for small deformation ($X/R<1$). Thus, we have%
\begin{equation}
U_{E}\sim ERY^{2} \label{UE}%
\end{equation}
The energy conservation, $MV^{2}\sim U_{E}$, allows us to obtain an estimate
for the contact time and a velocity-radius relation,%
\begin{align}
\tau &  \sim\tau_{c}\label{tE}\\
X/R  &  \sim V/V_{c} \label{xE}%
\end{align}

In the pancake regime at high velocities, where $X>R>Y$, the maximal energy
becomes $U_{P}\sim E\left(  (R+X)/R\right)  ^{2}R^{3}$ (the ideal rubber
deformation energy), or a linear spring energy%
\[
U_{P}\sim ERX^{2}%
\]
Noting that, in this case, the contact time might well be estimated not by
$Y/V$ but by $X/V$ due to an analogy with a spring system (physically, we can
imagine that after a strong impact the vertical velocity is immediately
redirected toward the horizontal direction), we obtain the same scaling
relations with the ellipsoid case (Eqs. (\ref{tE}) and (\ref{xE})).

The transition between the Hertz and the ellipsoid regimes is given by the
condition, $U_{E}\sim U_{H}$, which implies $Y\lesssim R$. In terms of
velocity this is expressed as $V\sim V_{c}$ (compare Eqs. (\ref{tH}) and
(\ref{tE})). On the other hand, the transition between the ellipsoid and the
pancake regimes should be marked by $X>Y$, or $V>V_{c}$. Since the transition
from Hertz to ellipsoid regimes and that from ellipsoid to pancake regimes are
predicted to occur at the same velocity at the level of scaling laws, the
ellipsoid regime is not expected to manifest itself clearly.

We summarize the above theoretical predictions which agree with experimental
observations: (1) $X$ scales as $X/R\sim(V/V_{c})^{8/5}$ for small velocities
but as $X/R\sim V/V_{c}$ for large velocities. (2) The contact time $\tau$ is
constant for large impact velocities but below a certain velocity it deviates
from the plateau value and increases with velocity decrease following a
scaling law, $\tau\sim V^{-1/5}$. Typical plateau value $\tau_{c}$ for our
samples is of the order of a few ms. It increases with $R$ while decreases
with $E$. (3) Both $X$ and $\tau$ can be divided in the two regimes by a
transition velocity $V_{c}$. Typical value $V_{c}$ is of the order of a few
m/s and it increases with $E$. (4) Data should be well characterized by
reducing size, time, and velocity variables by scales $R$, $\tau_{c}$, and
$V_{c}$, respectively.

In our Hertz regime, we assumed that the shape is a sphere cut out by a plane
and, in addition, we required Eq. (\ref{eqA}), which is natural at least in
the case of water drops where the surface energy is dominant \cite{MP,PGG}. As
a result, we have a global strain $X/R$ distributed over the whole volume
$R^{3}$, in addition to the original local distribution $Y/a$ over $a^{3}$.
The former extra energy is far smaller than the latter with a small ratio
$\simeq(a/R)^{3}$; this correction does not change the results of our previous analysis.

\section{Conclusion}

Experimentally, the scaling relation $X\sim V$ for large velocities is shown
and this relation seems to change into another scaling relation with a
stronger power below $V\simeq V_{c}$. Another scaling relation $\tau\sim
V^{-1/5}$ for small velocities is also shown while, for larger velocities
($V\gtrsim V_{c}$), $\tau$ seems to approach a plateau value. These behaviors
of $\tau$ and $X$ can be explained by the theory which starts from
experimentally observed shapes of balls at the maximal deformation and
employes the energy conservation. This theory also leads to the prediction
$X\sim V^{8/5}$ for small velocities, which seems consistent with experimental data.

\section{Discussion}

Experimentally it is difficult to observe a wide plateau region of the contact
time predicted by the pancake form and only an asymptotic behavior (or the
ellipsoid behavior) to this limiting regime is observed. This is because of
(1) it is practically difficult to achieve such high impact velocities by the
present experimental setup and (2) it is inherently impossible to exceed a
certain high impact velocity above which the impact causes irreversible
damages to gel balls.

The effect of gravity can be another source of increase in contact time with
decrease in impact velocity as might be the case in some bouncing water drops
\cite{KO}. In the present case, however, this possibility seems to be
excluded; the ratio $MgY/U_{H}\sim\rho gR(R/Y)^{3/2}/E$ suggests that the
gravitational energy $MgY$ becomes important only for $Y\lesssim0.04R$, which
is outside of our experimental region. Furthermore, if we define the length
$l_{g}$ by $\rho gl_{g}\sim E$, which is a counterpart of the capillary
length, this is about $1$m and is well beyond the characteristic length scale
$R$ (of the order of cm). This also suggests a weak gravity effect. These
arguments, in turn, suggest that the gravity may play a role for very small
deformations, which are not studied here.

The effect of viscosity is possibly more important and will be discussed in
detail elsewhere. In fact, if we closely look at Fig. \ref{f4} we find that
the behavior of the softest sample with the highest dissipation seems to
deviate slightly from the others (if we removed these points from these plots,
the data collapse would be much better). This suggests a possibility that
viscous effects become important already for the softest sample.

The characteristic velocity $V_{c}$ that emerged as the result of the static
energy evaluation turned out to coincide with the velocity of the shear wave
propagation, at the level of scaling laws. To understand this meaning, a
systematic study on the static large deformations of soft gel balls would be
helpful, and thus such measurements are now under study.

\section{Acknowledgements}

Y. T. thanks Yuji Kozono and Yoshi Miyamoto for their instruction in the
rheological measurements. Y. T. also thanks Takanobu Sato for his cooperation
at the early stage of this study. Y. T. and Y. Y. thank Mitsugu Matsushita for
his encouragement. K. O. benefited a lot from David Qu\'{e}r\'{e} through
useful discussions. K. O. is very grateful to P.-G. de Gennes for his third
stay in Paris financially supported by Coll\`{e}ge de France. K. O. also
appreciates Christophe Clanet and Fr\'{e}d\'{e}ric Chevy for discussions. This
work is also supported by an internal grant of Ochanomizu University.

\end{document}